\documentclass[journal]{IEEEtran}
\usepackage{cite}

\ifCLASSINFOpdf
   \usepackage[pdftex]{graphicx}
\else
\fi
\usepackage{amsmath}
\usepackage{amssymb}
\ifCLASSOPTIONcompsoc
  \usepackage[caption=false,font=normalsize,labelfont=sf,textfont=sf]{subfig}
\else
  \usepackage[caption=false,font=footnotesize]{subfig}
\fi

\usepackage{algorithmicx}
\usepackage[ruled]{algorithm}
\usepackage{algpseudocode}
\usepackage{algpascal}
\usepackage{algc}

\newcommand\ASTART{\bigskip\noindent\begin{minipage}[b]{0.5\linewidth}}
	
	\newcommand\AENDSKIP{\end{minipage}\bigskip}
\newcommand\AEND{\end{minipage}}

\usepackage{color}
\usepackage[dvipsnames]{xcolor}

\begin{document}
%

\title{A Novel Dynamic Peer-to-Peer Clustering Algorithm and Its Application to Aggregate Energy Storage Systems}

%
%

\author{Runfan Zhang,
        Branislav Hredzak,~\IEEEmembership{Senior Member,~IEEE}
\thanks{Runfan Zhang and Branislav Hredzak are with the School of Electrical Engineering and Telecommunications, University of New South Wales, Sydney, NSW 2052 Australia (email: runfan.zhang@student.unsw.edu.au; b.hredzak@unsw.edu.au;).}
}

%
%

%

\maketitle

\begin{abstract}
The proposed distributed dynamic clustering algorithm enables to group agents based on their pre-selected feature states. The clusters are determined by comparing the distance of the agents' current feature states with average estimates of the states in all clusters. The algorithm also provides average estimates of pre-selected auxiliary states that can be utilized for control purposes. Two example applications of the algorithm are introduced. In the first application, the algorithm is applied to a microgrid with distributed batteries that are controlled to achieve a common state of charge within a group. However, a random selection of the batteries' groups results in additional power losses during operation. The algorithm reduces the power losses by clustering the batteries based on the selected feature states: local loads and battery capacities, while the state of charges and output voltages are selected as auxiliary states for control purposes. In the second application, the algorithm is used to form a virtual energy storage from batteries distributed in a microgrid.
\end{abstract}


%
\IEEEpeerreviewmaketitle

\section{Introduction}
%
%
%
%
\IEEEPARstart{C}{lustering} is a division of data into groups with similar properties \cite{Clustring_Df}. Various centralized clustering algorithms have been researched, such as k-means \cite{KmeanC}, hierarchical clustering \cite{HierarchicalC}, self-organization map \cite{SOM} or expectation maximization clustering algorithms \cite{EMA}. Besides, distributed clustering algorithm have been proposed for distributed data environment \cite{DisCluster}. However, although the data is distributed, these algorithms require a main site for global clustering, with each data storage communicating with the main site through a centralized network \cite{DisClstrRv}. In addition, all before-mentioned algorithms are clustering based on previously collected data and not on real-time system states. Therefore, these algorithms are inadequate for clustering multi-agent based systems communicating via a distributed network. 

Microgrids are small-scale power networks that supply local loads in small geographical areas. The microgrid enables penetration of renewable energy sources, such as photovoltaic (PV) and wind generation, and energy storage \cite{CoopMG_bk}. To control the microgrid, the traditional hierarchical power systems control architecture can be applied \cite{BMG_RV}: primary, secondary and tertiary control levels. The primary control level applies droop control to provide proportional power sharing. Then, the secondary control eliminates the voltage and/or frequency deviations caused by the droop control and, at the same time, provides accurate power sharing. Power flow objectives objectives are realized at the tertiary control level. 

To improve the power quality and reliability in microgrids, a state of charge (SoC) balancing control strategy at the secondary control level was proposed for distributed energy storages (ESs) \cite{BMG_RV}. Not only the strategy maintains the SoCs of all ESs at the same level, but it can also provide a proportional power sharing between ESs after a balanced SoC is achieved. However, for a microgrid with a large number of heterogeneous ESs and different local loads, maintaining all ESs SoCs at the same level is not always the best option. This is because during and after the SoC balancing, the power network line currents inevitably result in power losses.

Literature on application of clustering algorithms in microgrids can be broadly divided into three categories. In the first category, power systems are clustered based on previous data profiles, such as 24 hour data profiles of loads and renewable energy generation \cite{clstrLoad,clstrMG2,ClstringWindfarm}. The purpose of clustering is to achieve economical operation. The second category is based on stochastic modeling utilizing a one year data profile (365 days) and dividing the data into different probabilistic scenarios to deal with uncertainties \cite{clsterTime,ClstrsStochastic}. The last one defines a set of power systems in a geographical area without using any clustering algorithm \cite{clsterMG,VBES,clusterPS}. Besides, clustering methods were used for partitioning a central network based on voltage source converter sensitivity matrices, voltage influence factors and merge matrices \cite{Netclustering}, and by finding a critical load \cite{CLoadClstr}. However, all above clustering algorithms require past data and are run off-line, i.e. the clustering is determined before the system operation.

Motivated by the above discussion, this paper proposes a novel, fully distributed dynamic on-line clustering algorithm based on dynamic states of multi-agents communicating through a distributed communication network. The clustering algorithm dynamically aggregated ESs into clusters with selected similar features and maintains a balanced SoC within each cluster. The proposed algorithm is applied (i) to mitigate power losses associated with the traditional SoC balancing of fixed ES clusters distributed in a microgrid and (ii) to form a virtual energy storage . 
The salient novel features of this paper are:
\begin{enumerate}
	\item The clustering algorithm is fully distributed and can be implemented with any existed distributed control method.
	\item The clustering is done on-line, i.e. the clustering can dynamically adjust to variation in multi-agent states and does not require previously collected data. 
	\item The proposed algorithm is applied to a microgrid (i) to reduce power losses associated with SoC balancing of batteries in fixed clusters and (ii) to form a virtual energy storage.
	\item The clustering algorithm is robust to multi-agents' state variation. As a result, frequent cluster changes, caused by fast state variations, can be eliminated.
\end{enumerate}

The rest of this paper is organized as follows. Section \ref{sec_DCA} presents the proposed algorithm. Implementation of the algorithm to a microgrid with distributed batteries is discussed in Section \ref{sec_ClstrMG}. Section \ref{sec_Conclusion} concludes the paper.

\section{Distributed Clustering Algorithm}\label{sec_DCA}
This section introduces the proposed distributed clustering algorithm for multi-agents. The clustering problem can be defined as: cluster $N$ agents into $M$ groups based on some selected features propagated through a distributed network. Also assume that the estimations of average states inside of a cluster can be obtained by each agent through the distributed network.

Next, the concept of a distributed network is introduced and then the proposed algorithm is presented.
\subsection{Distributed Communication Network}
Multi-agents communicate with neighbours through a network represented by a sparse graph $\mathcal{G}\left(\mathcal{V},\mathcal{E}\right)$, with nodes $\mathcal{V}=\{1,\cdots,N\}$ and edges $\mathcal{E}$ \cite{coopCtrl}. Each graph node represents an agent. Elements of $\mathcal{E}$ are denoted as $\left( {i,\;j} \right)$, where $\left( {i,\;j} \right) \in \mathcal{E}$ if there is a link allowing information to flow from node $i$ to node $j$. The neighbours of node $i$ are given by $\mathcal{N}_{i}$, where $j \in {{\mathcal{N}}_i}$, if $\left( {i,\;j} \right) \in \mathcal{E}$. The graph adjacency matrix is given by
\begin{equation}
\label{Adj}
{\mathcal{A}} = \left[ {{a_{ij}}} \right] \in {\mathbb{R} ^{N \times N}}, {a_{ij}} = \left\{ {\begin{array}{*{20}{c}}
	{\alpha,\;\left( {j,\;i} \right) \in \mathcal{E} }\\
	{0,\;otherwise}
	\end{array}}, \right.
\end{equation}
where $\alpha$ is the coupling strength. Then, the graph Laplacian matrix is $L=\mathcal{D}-\mathcal{A}$, where $\mathcal{D}=diag\{d_i\}$, and ${d_i} = \sum\nolimits_{j = 1}^N {a_{ij}}$ is the in-degree of the communication network.
\subsection{Dynamic Clustering Algorithm}
\begin{figure}[!t]
	\centering
	\includegraphics[width=3.5in]{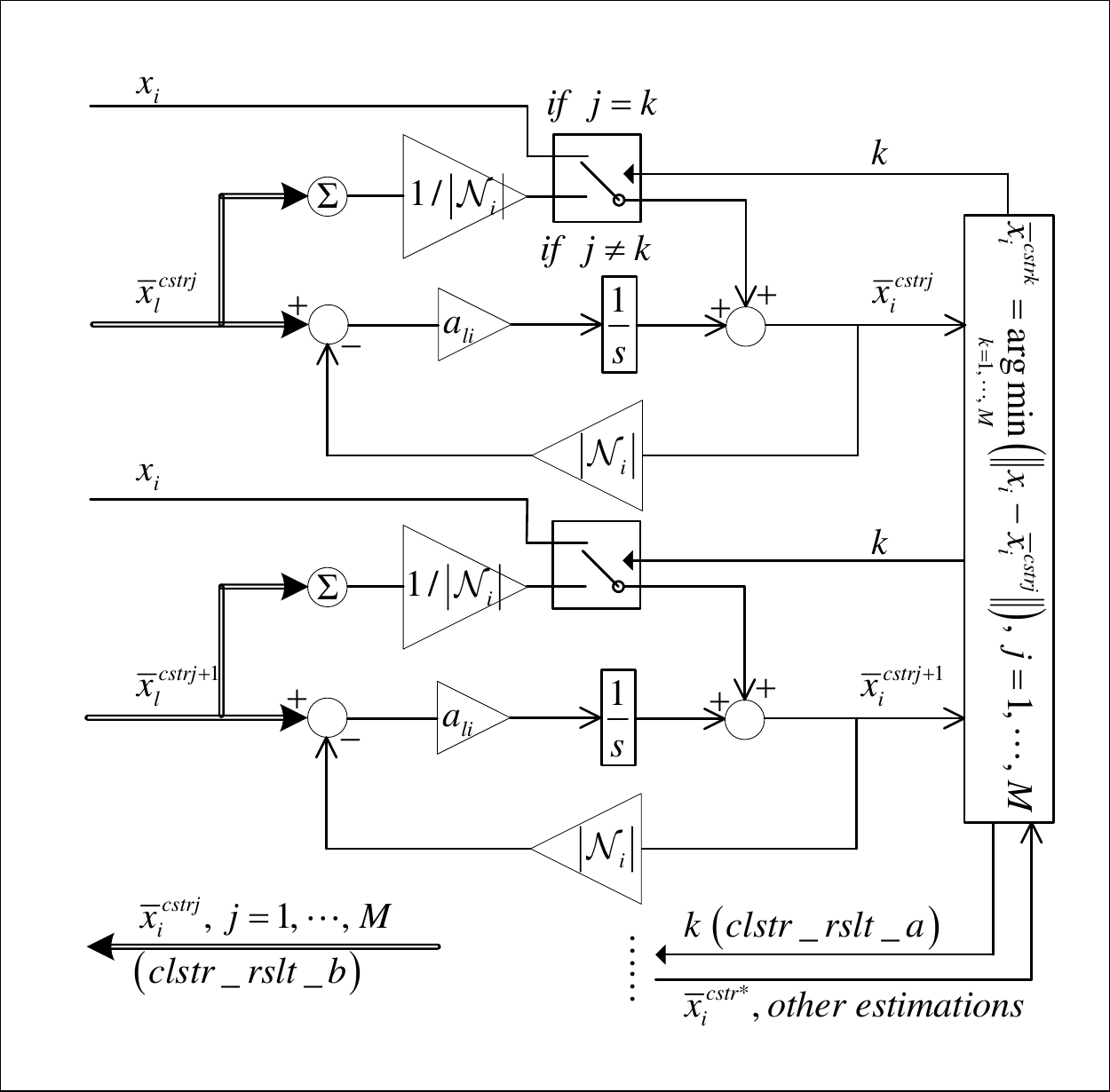}
	\caption{Distributed clustering algorithm for the $i$-th agent. The double line arrows represent distributed communication.}
	\label{fig_Agrm}
\end{figure}
\begin{figure}[!t]
	\centering
	\includegraphics[width=3in]{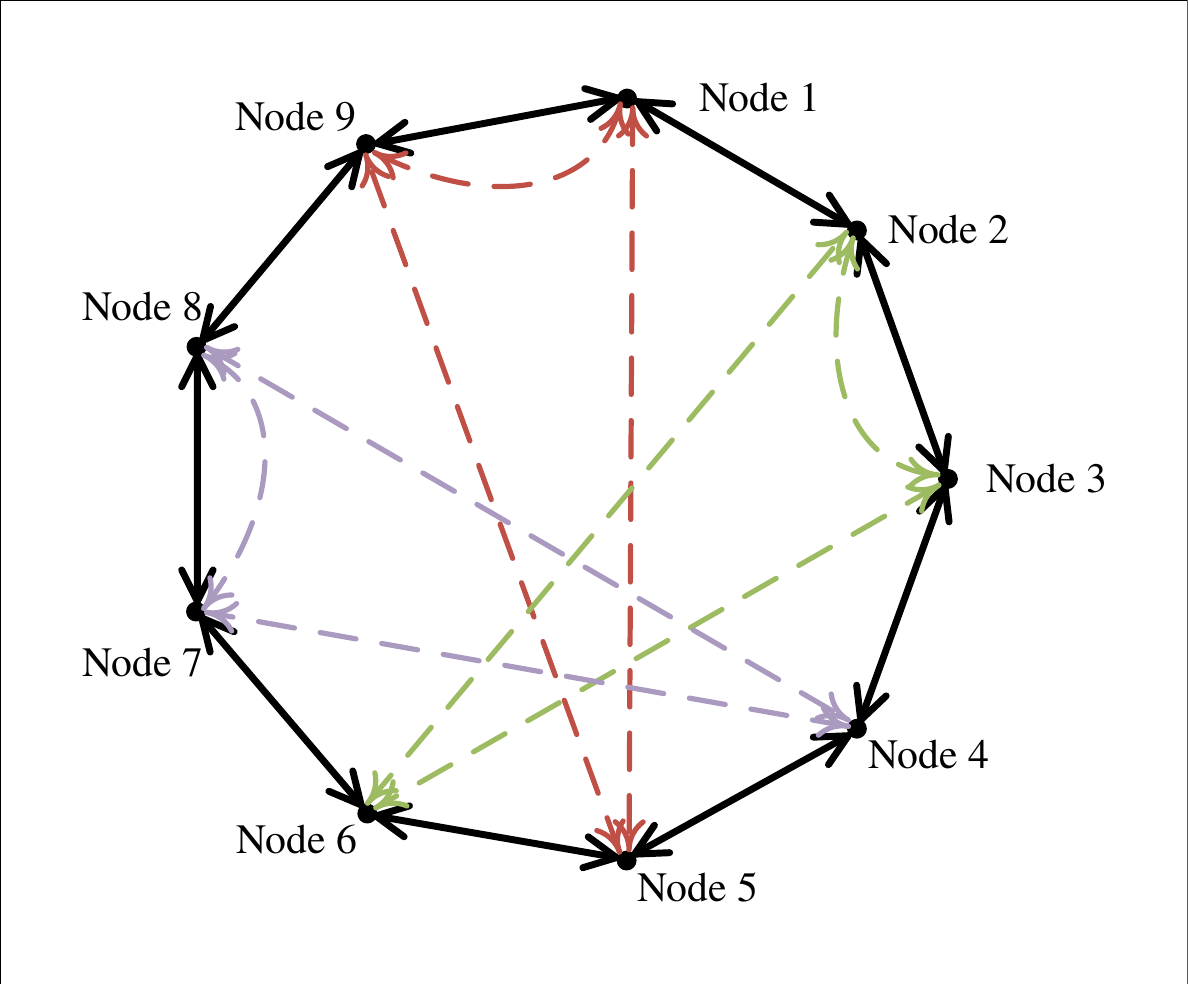}
	\caption{Illustration of the proposed clustering algorithm. Nine agent nodes are clustered into three groups. The black lines represent the distributed communication network between agent nodes. The colored dash lines indicate virtual communications within a cluster. One color represents one cluster.}
	\label{fig_AgrmSMP}
\end{figure}
The feature states of an agent, used to determine the clusters, are defined as ${x_i} \in {\Re ^n}$. The proposed algorithm dynamically clusters $N$ agents into $M$ clusters based on the feature states. Estimates of the average of the states ${x_i}$ in the $j$-th cluster are $\bar x_i^{cstrj}$, where $i = 1,\; \cdots ,\;N$ and $j = 1,\; \cdots ,\;M$. Each $i$-th agent has continuous access to the latest average states $\bar x_i^{cstrj}$ of all clusters through the communication network.

First, select (arbitrary) initial estimates of $\bar x_i^{cstrj}$. Then, in order to decide/update to which cluster the $i$-th agent belongs, its current state $x_i$ is compared with all current estimates of $\bar x_i^{cstrj}, j = 1,\; \cdots ,\;M$. The $k$-th estimate $\bar x_i^{cstrk}$ which has the smallest distance from the current state ${x_i}$, as given by \eqref{eq_DCA}, denotes that the $i$-th agent is allocated to the $k$-th cluster. Once all clusters are determined, each $i$-th agent estimates the average of the states of the $k$-th cluster to which it belongs based on its current state ${x_i}$ and the $k$-th cluster agents' estimates passed through the network, as given by \eqref{eq_avg_xk}. Neighbours' estimates that do not belong to the $k$-th cluster, $j \in \left\{ {1,\; \cdots ,\;M} \right\},j \ne k$, are passed through, as given by \eqref{eq_avg_xj}.

\begin{align}
&\dot {\bar {x}}_i^{cstrk} = {{\dot x}_i} + \sum\limits_{l \in {\mathcal N_i}} {{a_{il}}\left( {\bar x_l^{cstrk} - \bar x_i^{cstrk}} \right)}, \label{eq_avg_xk} \\
&\dot {\bar {x}}_i^{cstrj} = \frac{1}{\left| {\mathcal N_i} \right|}\sum\limits_{l \in {\mathcal N_i}} {\dot {\bar {x}}_l^{cstrj}}  + \sum\limits_{l \in {\mathcal N_i}} {{a_{il}}\left( {\bar x_l^{cstrj} - \bar x_i^{cstrj}} \right)} ,\;j \ne k,\label{eq_avg_xj}
\end{align}
where ${\left| {\mathcal N_i} \right|}$ denotes the number of neighbours of the $i$-th agent. 
 
\begin{figure*}[!t]
	\normalsize
\begin{equation}\label{eq_DCA}
{x_i} \in \left\{ {k \text{-th cluster}\left| {\bar x_i^{cstrk} = \mathop {\arg \min }\limits_{k \in \left\{ {1,\; \cdots ,\;M} \right\}} \left( {\left\| {{x_i} - \bar x_i^{cstrj}} \right\|} \right),\;j = 1,\; \cdots ,\;M} \right.} \right\}.
\end{equation}
\hrulefill
\vspace*{4pt}
\end{figure*}

Each $i$-th agent has always access to two results: 
\begin{enumerate}
	\item $clstr\_rslt\_a$: information to which $k$-th cluster it belongs;
	\item $clstr\_rslt\_b$: estimates of the average of the feature states $\bar x_i^{cstrj}$ in the $j$-th cluster, $j = 1,\; \cdots ,\;M$.
\end{enumerate} 

The algorithm is illustrated in Fig. \ref{fig_Agrm} and can be summarized as \textbf{Algorithm \ref{alg_ddyc}}.

\alglanguage{pseudocode}
\begin{algorithm}[H]
	\caption{Distributed Dynamic Clustering}\label{alg_ddyc}
	\algblock{Each}{endEach}
	\algnewcommand\algorithmicEach{\textbf{each}}
	\algnewcommand\algorithmicEEach{\textbf{end\ each}}
	\algrenewtext{Each}[1]{\algorithmicEach\ #1}
	\algrenewtext{endEach}{\algorithmicEEach}%
	\begin{algorithmic}[1]
		\State design $\mathcal{G}\left(\mathcal{V},\mathcal{E}\right)$ and $M$
		\State initialize $\bar x_i^{cstrj}, j = 1,\; \cdots ,\;M$
		\Each {$i$-th agent at time $t$} 
			\State measure $x_i$
			\State receive ${\bar {x}}_l^{cstrj}, {l \in {\mathcal N_i}}, j = 1,\; \cdots ,\;M$ from neighbours
			\State $k \gets$ label of the cluster with the smallest distance between $x_i$ and $\bar x_i^{cstrj}, j = 1,\; \cdots ,\;M$, where $k \in \left\{ {1,\; \cdots ,\;M} \right\}$
			\If {$j=k$}
				\State Eq. \eqref{eq_avg_xk}
			\Else
				\State Eq. \eqref{eq_avg_xj}
			\EndIf
			\State send $\bar x_i^{cstrj}, j = 1,\; \cdots ,\;M$ to neighbours
		\endEach
		\State Each local agent $i$ implements the same algorithm.
	\end{algorithmic}
\end{algorithm}

\textit{Example}: To further clarify the algorithm, consider the communication network structure in Fig. \ref{fig_AgrmSMP}. In the figure, nine agent nodes are clustered into three groups. In each cluster, the average estimations are calculated based on the agents in the group, i.e. only the agents in the same cluster communicate with each other to obtain the average estimates through virtual communication lines shown as colored dash lines. The agents which are not in the cluster simply pass trough estimates of the other clusters to its neighbours. Assume the red dashed line connecting nodes 1, 5 and 9 denote the $k$-th cluster. Nodes 1, 5, and 9 obtain the average estimate of the cluster $\bar x_i^{cstrk}$  using \eqref{eq_avg_xk}. However, for the other two clusters' estimates $\bar x_i^{cstrj}, j \ne k$, nodes 1, 5, and 9 apply \eqref{eq_avg_xj} to estimate the averages. Similarly, the other six agents ($i \ne 1,\;5,\; \text{and}\;9$) obtain the average estimates of the red line cluster $k$,  $\bar x_i^{cstrk}, i \ne 1,\;5,\; \text{and}\;9$, by applying \eqref{eq_avg_xj}. As a result, the nodes 1, 5, and 9 in the red line cluster $k$ directly communicate with each other through the virtual red color line links. The other agents just transmit the estimations $\bar x_i^{cstrk}, i=1,\;5,\; \text{and}\;9$.

\subsection{Utilization of Clusters}
\begin{figure}[!t]
	\centering
	\includegraphics[width=3.5in]{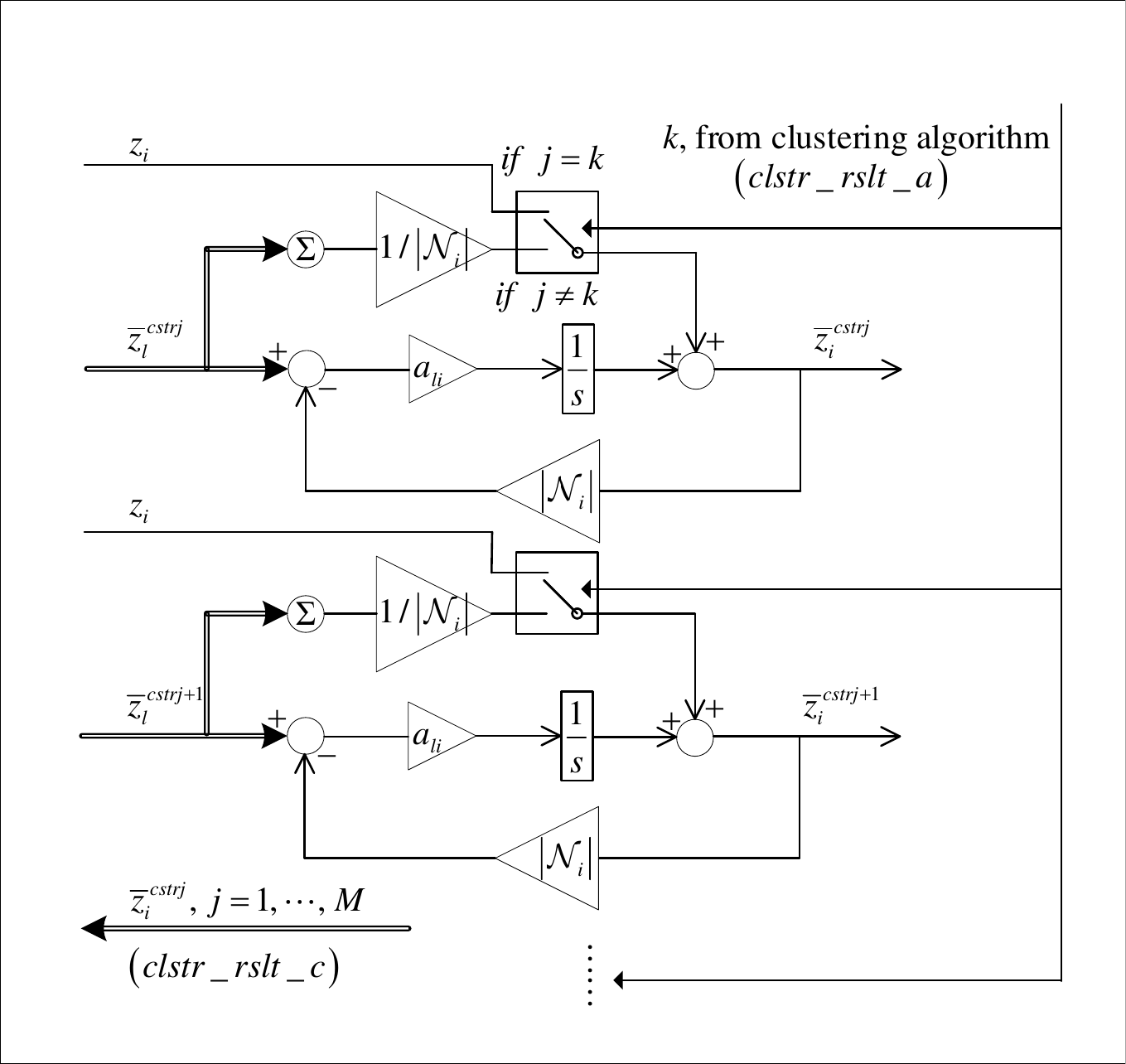}
	\caption{The average estimations of the auxiliary state $z_i$ for the $i$-th agent. The double line arrows represent distributed communication.}
	\label{fig_Agrmz}
\end{figure}
 In some cases, other states than the feature states may be required for control, optimization and analysis within the cluster or between the clusters. These states are defined as the auxiliary states, ${z_i} \in {\Re ^m}$, and are only used to calculate average of the states within a cluster. For example, in a microgrid, local loads and battery sizes are selected as feature states to determine clusters, whereas the SoCs are selected as auxiliary states for the average estimates of SoCs and the SoC balancing control within a cluster.

Hence, similar to feature states, the estimates of the auxiliary states ${z_i}$ are $\bar z_i^{cstrj}$, where $i = 1,\; \cdots ,\;N$ and $j = 1,\; \cdots ,\;M$. The estimates of the average states of the $i$-th agent in the $j$-th cluster can be obtained using the $clstr\_rslt\_a$ result and following the same procedure as for the estimates $\bar x_i^{cstrj}$,
\begin{align}
&\dot {\bar {z}}_i^{cstrk} = {{\dot z}_i} + \sum\limits_{l \in {\mathcal N_i}} {{a_{il}}\left( {\bar z_l^{cstrk} - \bar z_i^{cstrk}} \right)}, \label{eq_avg_zk}\\
&\dot {\bar {z}}_i^{cstrj} = \frac{1}{\left| {\mathcal N_i} \right|}\sum\limits_{l \in {\mathcal N_i}} {\dot {\bar {z}}_l^{cstrj}}  + \sum\limits_{l \in {N_i}} {{a_{il}}\left( {\bar z_l^{cstrj} - \bar z_i^{cstrj}} \right)} ,\;j \ne k. \label{eq_avg_zj}
\end{align}
Hence, each $i$-th agent has always access to $clstr\_rslt\_c$ result: estimates of the average of the auxiliary states $\bar z_i^{cstrj}, j = 1,\; \cdots ,\;M$. The algorithm is illustrated in Fig. \ref{fig_Agrmz}. 

Including the auxiliary states into the \textbf{Algorithm \ref{alg_ddyc}} gives  \textbf{Algorithm \ref{alg_ddycAE}}.
 \begin{algorithm}[H]
	\caption{Distributed Dynamic Clustering with Utilization of Clusters}\label{alg_ddycAE}
	\algblock{Each}{endEach}
	\algnewcommand\algorithmicEach{\textbf{each}}
	\algnewcommand\algorithmicEEach{\textbf{end\ each}}
	\algrenewtext{Each}[1]{\algorithmicEach\ #1}
	\algrenewtext{endEach}{\algorithmicEEach}%
	\begin{algorithmic}[1]
		\State design $\mathcal{G}\left(\mathcal{V},\mathcal{E}\right)$ and $M$
		\State initialize $\bar x_i^{cstrj}, j = 1,\; \cdots ,\;M$
		\Each {$i$-th agent at time $t$}
		\State measure $x_i$ and $z_i$
		\State receive ${\bar {x}}_l^{cstrj}, {l \in {\mathcal N_i}}, j = 1,\; \cdots ,\;M$ from neighbours
		\State receive ${\bar {z}}_l^{cstrj}, {l \in {\mathcal N_i}}, j = 1,\; \cdots ,\;M$ from neighbours
		\State $k \gets$ label of the cluster with the smallest distance between $x_i$ and $\bar x_i^{cstrj}, j = 1,\; \cdots ,\;M$, where $k \in \left\{ {1,\; \cdots ,\;M} \right\}$
		\If {$j=k$}
		\State Eq. \eqref{eq_avg_xk} and \eqref{eq_avg_zk}
		\Else
		\State Eq. \eqref{eq_avg_xj} and \eqref{eq_avg_zj}
		\EndIf
		\State send $\bar x_i^{cstrj}, j = 1,\; \cdots ,\;M$ to neighbours
		\State send $\bar z_i^{cstrj}, j = 1,\; \cdots ,\;M$ to neighbours
		\endEach
		\State Each local agent $i$ implements the same algorithm.
	\end{algorithmic}
\end{algorithm}

\section{Applications of Algorithm to Cluster Batteries in Microgrid}\label{sec_ClstrMG}

This section introduces two example applications of the proposed algorithm to cluster batteries distributed in a microgrid in order (i) to reduce the power losses and (ii) to form a virtual energy storage.

A battery energy storage system (BESS) includes a battery, a bidirectional DC-DC converter interfacing the battery with the microgrid and a local load. 
\subsection{Power Loss Reduction}
Balancing SoC of batteries distributed in a microgrid can improve stability and reliability of the power system by making full use of energy stored in batteries and lengthening the battery lifetime \cite{BMG_RV}. However, if the batteries are randomly clustered into fixed groups, it can result in power losses even though their SOCs are balanced. The additional power losses are caused by different local loads and different battery capacities, as explained next. 


\subsubsection{Local Loads} Assume that all batteries have the same capacity. The power losses, after a balanced SoC has been achieved, can be reduced by clustering the battery systems with similar local loads. This is because if each battery in a cluster has the same load, there will be no line currents flowing between the batteries.


\subsubsection{Battery Capacity} Assume that all batteries have equal local loads. After a balanced SoC has been reached, larger capacity batteries have to provide more power than smaller capacity batteries to maintain the balanced SoC. This results in additional line currents and hence power losses. Therefore, clustering batteries with similar capacity can reduce the power losses.

Based on the above discussion, two feature states: local load and battery capacity are selected to cluster the batteries using the proposed distributed clustering algorithm and hence minimize the power losses.

\subsection{Virtual Energy Storage}
In this application, the batteries are clustered into virtual energy storage(s) to sell power to the main grid or to absorb the power from renewable energy sources. 

It is assumed that the utility company provides required energy storage capacity (virtual energy storage) and its price to be formed from distributed BESSs. At the same time, all prosumers decide how much of their battery capacity will be made available to form the virtual energy storage and at what price. Then, the utility company pins (broadcasts) its virtual energy storage requirements to at least one prosumer in an area where prosumers can exchange information through a neighbour-to-neighbour distributed communication network. The requirements are the initial values for the clusters. Two feature states, price and battery capacity, are applied in the algorithm. The algorithm finds the clusters of batteries that are close to the feature states' average estimations. All prosumers in a cluster can sell their available battery capacity at the average price. Subsequently, the company receives all information from the pining prosumers and selects clusters with sufficient energy capacity and at acceptable price to form the virtual energy storage. The SoC balancing control algorithm maintains the same SoC level in all batteries in the virtual energy storage by ensuring that batteries provide power in proportion to their own energy capacity.

It should be noted that the selected feature states are not restricted to the price and energy capacity only, additional feature states, such as distance or geographical location, can be added to form virtual energy storage(s).

\section{Conclusion}\label{sec_Conclusion}
This paper proposed a distributed dynamic clustering algorithm for clustering agents based on pre-selected feature states. The clustering algorithm was applied (i) to cluster batteries with a similar load and energy capacity to reduce the power losses and (ii) to form a virtual energy storage based on price and energy capacity. The proposed algorithm can be applied to any other systems which require dynamic, on-line clustering based on pre-selected feature states, while using auxiliary states for any required control, optimization or analysis purposes.  


%



\section*{Acknowledgment}

R. Zhang would like to gratefully thank for a scholarship from the China Scholarship Council.

\ifCLASSOPTIONcaptionsoff
  \newpage
\fi



\bibliographystyle{IEEEtran}
\end{document}